\documentclass[conference]{IEEEtran}
\IEEEoverridecommandlockouts

\usepackage[margin=1.6cm]{geometry}
\usepackage{graphicx}
\usepackage{wrapfig}
\usepackage{lscape}
\usepackage{rotating}
\usepackage{seqsplit}
\usepackage{subfig}

\usepackage{algorithmic,amsfonts,amsmath,amsthm,amssymb,tikz,float,url,palatino,bbm}
\usepackage[utf8]{inputenc}
\usepackage[T1]{fontenc}

\usepackage{float}
\usepackage{xcolor}
\usepackage{booktabs}
\usepackage{combelow}
\usepackage{multirow}
\usepackage{colortbl}

\usepackage{soul}

\def\BibTeX{{\rm B\kern-.05em{\sc i\kern-.025em b}\kern-.08em
    T\kern-.1667em\lower.7ex\hbox{E}\kern-.125emX}}

\renewcommand{\phi}{\varphi}

\usepackage[ruled]{algorithm2e}

\title{Banked Memories for Soft SIMT Processors}

\author{\IEEEauthorblockN{Martin Langhammer\IEEEauthorrefmark{1}\IEEEauthorrefmark{2} and
George A.~Constantinides\IEEEauthorrefmark{2}}
\IEEEauthorblockA{\IEEEauthorrefmark{1} Altera Corporation\\ \IEEEauthorrefmark{2} Electrical and Electronic Engineering, Imperial College London\\
Email: martin.langhammer@altera.com,
g.constantinides@imperial.ac.uk}}

\graphicspath{{Figures/}}
\begin{document}
\maketitle

\begin{abstract}
Recent advances in soft GPGPU architectures have shown that a small (<10K LUT), high performance (770 MHz) processor is possible in modern FPGAs. In this paper we architect and evaluate soft SIMT processor banked memories, which can support high bandwidth (up to 16 ports) while maintaining high speed (over 770 MHz). We compare 9 different memory architectures, including simpler multi-port memories, and run a total of 51 benchmarks (different combinations of algorithms, data sizes and processor memories) to develop a comprehensive set of data which will guide the reader in making an informed memory architecture decision for their application. Our benchmarks are comprised of matrix transpositions (memory intensive) and FFTs (split between memory accesses, floating point, and integer computations) to provide a balanced evaluation. We show that the simpler (but more memory block intensive) multi-port memories offer higher performance than the more architecturally complex banked memories for many applications, especially for smaller memories, but the effective footprint cost of the multi-port memories quickly becomes prohibitive as dataset sizes increase. Our banked memory implementation results - high bandwidth, high Fmax, and high density - can be used for other FPGA applications as well, such as HLS (High Level Synthesis).
\end{abstract}

\section{Introduction}

At first glance, the FPGA would seem to be well suited to the construction of SIMT (Single Instruction Multiple Thread) processors such as GPGPUs. There is a high density of embedded arithmetic (current FPGAs~\cite{AgilexAGF027} have similar levels of IEEE754 floating point density -- 10 to 20 TFLOPS -- to current GPGPUs~\cite{A100}). The many SPs (scalar processors) that comprise the SMs (streaming multiprocessors) of GPGPU architectures are individually relatively simple multi-threaded processors. These consist of a large amount of registers (which can be easily mapped to the large number of embedded M20K memories in the FPGA) which are tightly bound to floating point and integer arithmetic units (again, which can be easily mapped to the large number of available DSP Blocks).

Earlier GPGPU architectures mapped to FPGA tended to be low performance (50MHz - 300MHz), and/or very large (60K - 300K LUTs)~\cite{guppy,FlexGrip,FlexGrip_Thesis,MIAOW,SCRATCH,FGPU,DOGPU}. This work builds on the eGPU project~\cite{FPL_eGPU}~\cite{ISFPGA_eGPU}, which is a area efficient (5K-12K ALMs), high performance GPGPU (771 MHz, limited by the speed of the DSP Blocks in floating point mode) on Intel Agilex-7 devices~\cite{Agilex}. In this paper, we describe a SIMT processor that ranges from 8K LUTs to 24K LUTs (the latter size includes a 16 bank memory of 448KB), which closes timing in excess of 771 MHz for an unconstrained compile. 

The peak performance of a SIMT processor is set by the arithmetic density (TFLOPs), but the sustained performance (which is often a small fraction of the peak) is usually limited by the memory bandwidth. Multi-port memory has a modest fixed bandwidth, a high memory cost, and a low control logic cost. Banked memory, commonly found in commercial GPUs~\cite{TeslaV100}~\cite{A100}, has a large {\em{potential}} (but highly variable) bandwidth, a modest memory cost, but a very high control cost and complexity.

Unlike a standard GPGPU which has multiple levels of memory hierarchy~\cite{TeslaV100,A100}, we use the shared memory for main data storage. Global reading and writing is in and out of the shared memory. This additional functionality of the shared memory in a soft GPGPU context necessitates a relatively large shared memory.

The goal of this paper is to look at the effect of memory architecture only. There are many known methods, such as other architecture enhancements and programming optimizations we can apply to improve performance, but we will not explore these in this paper, so that any differences in benchmark performance can be attributed to memory architecture alone.

We make the following contributions in this work:
\begin{itemize}
    \item We design and implement a scalable (in both memory size and number of banks) banked memory system, which closes timing in excess of 775 MHz in an unconstrained compile, and over 735 MHz in a tightly constrained compile.
    \item We describe FPGA design techniques, including resource depth matching, and carry-chain based arbitration, so that the reader can reproduce our results. 
    \item We benchmark several matrix transpose and FFT programs over 9 memory architectures (both multi-port and banked types, including different bank mapping methods), to build up a more complete picture of soft SIMT implementation tradeoffs.
    \item We benchmark our results using true footprint area, rather than simple resource count, and explain why this is important in the context of high-performance, tightly-packed designs. 
\end{itemize}

\section{Prior FPGA Banked Memory}

Most FPGA high-bandwidth memory (multiple read and write ports) published research implementations are multi-port ~\cite{LaForest}~\cite{LaForest_XOR}~\cite{Multiport_Lemieux}~\cite{BRAM_Efficient}, rather than banked architectures. Multi-port memories have much simpler arbitration mechanisms, but a very high memory cost because the data must be replicated. Banked memory use is infrequent, and usually associated with HLS (High Level Synthesis) tools~\cite{Zhou_banks,Chen_banks,EASY_banks}, as the access arbitration can often be simplified by static analysis of the compiled program. (While banked memory architectures are also part of many of the FPGA parallel processors~\cite{FGPU,FlexGrip}, little detail is reported in the soft processor work).

In an HLS design, the memory accesses can be profiled using several different methods, such as trace-driven address mining, which looks at the generated addresses~\cite{Zhou_banks,Chen_banks} or a formal verification of a static program analysis~\cite{EASY_banks}. Unique mappings can be used for HLS because there is typically only a single program running. In contrast, a processor will need to support user-generated programs, which will require a generic mapping algorithm. 

The application-specific mappings generated for the HLS designs often reduce the complexity of the crossbars, as not every thread will access every memory bank. For a 16-bank memory, the unoptimized HLS banked memory requires 11K to 24K ALMs~\cite{EASY_banks}, which is in the same range as our general-purpose core. The performance, however, is only around 100 MHz. This is partially because of the lower performance Cyclone V~\cite{CycloneV} device used - we might expect a higher performance Agilex~\cite{Agilex} device to achieve 50\% faster frequency.

\section{Banked Memory Architectures for Soft SIMT}
Resolving bank conflicts - or indeed avoiding them as much as possible - is one of the key challenges in using a banked memory. The eGPU originally avoided bank conflicts by using a multi-port memory, with 4 read ports, and 1 or 2 write ports, depending on the memory style selected. The write bandwidth was found to be a significant performance bottleneck~\cite{ISFPGA_eGPU}. One advantage of the multi-port memory, however, was deterministic access time. In addition to simplifying the programming model, this also simplified the architecture, contributing to the very high performance.

In a SIMT architecture, an instruction will typically execute all threads before starting the next instruction. In the case of the eGPU, an example program configured with 4096 threads will require 256 clocks (4096 threads/16 SPs) per instruction. (In Nvidia terms, the thread block is 4096 and the warp 16). For the purposes of this paper we will call the 16 threads issued per clock a memory {\em{operation}}, and each individual thread memory access a {\em{request}}. In other words, there are 16 memory requests per clock.

A banked memory introduces variable shared memory access times. If there are no bank conflicts, a read or write operation completes in a single clock cycle. At the other extreme, if all 16 SPs need to access the memory in only one of the banks, the read or write operation will take 16 cycles.

\begin{figure}
	\begin{center}
		\includegraphics[scale=0.56]{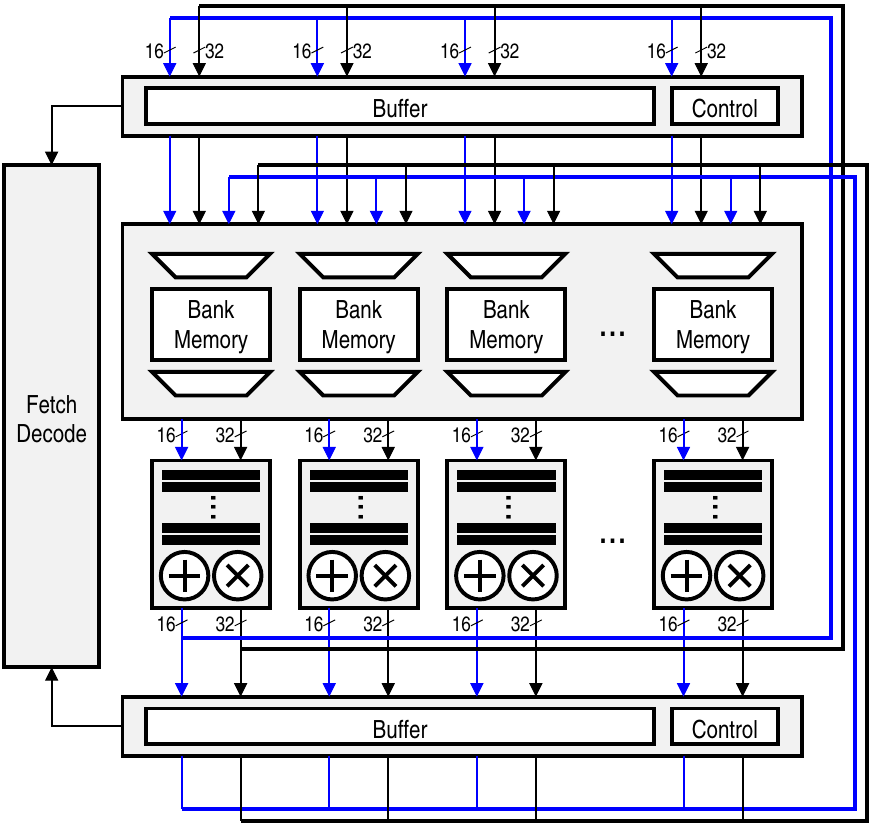}
		\caption{Top Level Architecture of SIMT Processor with Parallel Banked Memory}
		\label{fig:top_level} 
	\end{center}
\end{figure}

Figure~\ref{fig:top_level} shows the major components of our SIMT architecture. There are 16 parallel SPs, connected to a shared memory with 16 banks. This supports a massive $potential$ bandwidth of 16 ports; although in practice, it is often less. The shared memory block diagram contains a large number of mux structures, which we will see is a major contributor to the area (ALMs - which are the fracturable 6-input LUTs that make up the Intel FPGA soft logic) and routing congestion in this design. There are both read and write access controller blocks, which manage access into the shared memory. Both access controllers can stall or hold the instruction fetch and decode as well. Memory banks are 32 bits wide, and the index of each of the 16 memory banks is the lower 4 bits of the read or write address. We can modify this indexing, and we will examine the benchmark impact of this later in this paper. 

\begin{figure}
   \centering
   \includegraphics[scale=0.55]{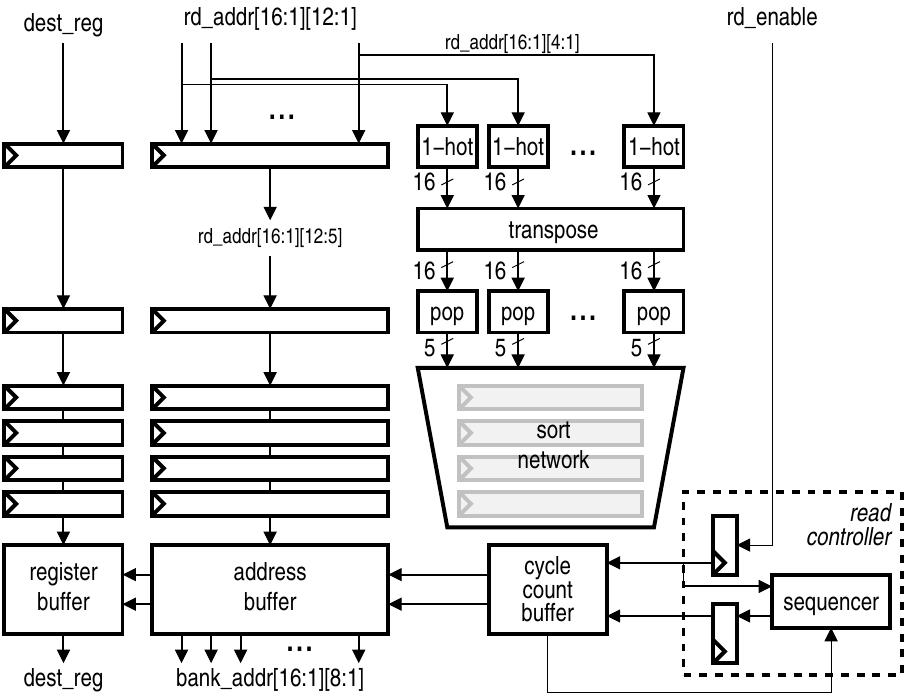}
   \caption{Read Access Control to Shared Memory}
  \label{fig:eGPU_read_ctl}
\end{figure}

\begin{figure}
    \centering
    \includegraphics[scale=0.55]{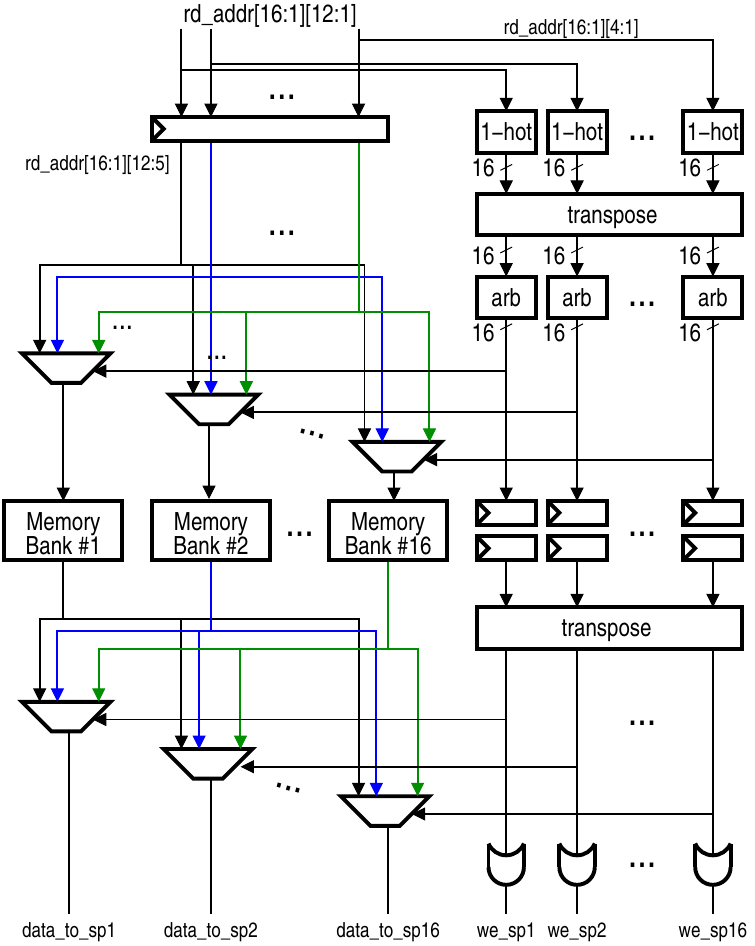}
   \caption{Read Arbitrate for Shared Memory}
   \label{fig:eGPU_read_arbitrate}
\end{figure}

\begin{figure*}
    \centering
    \includegraphics[scale=0.70]{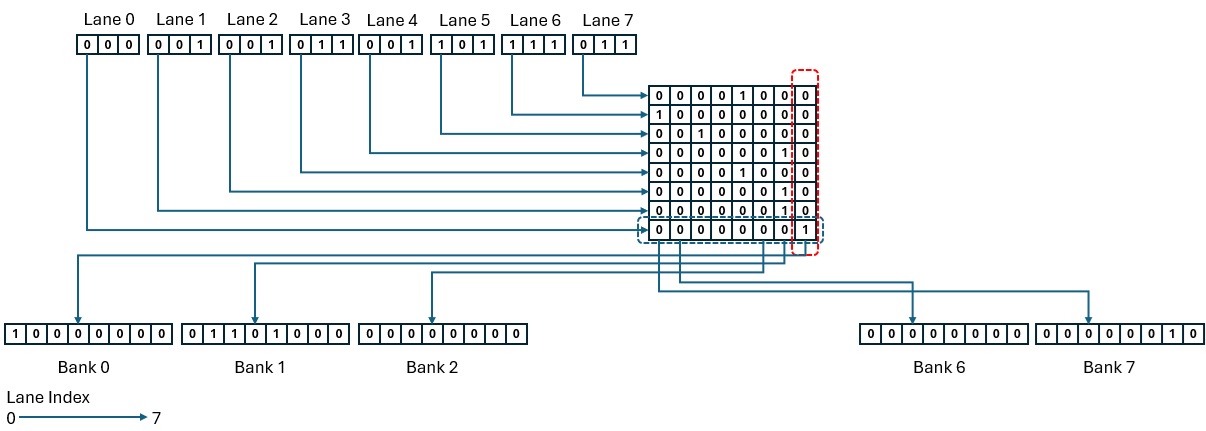}
   \caption{Bank Mapping}
   \label{fig:eGPU_bankmap}
\end{figure*}

\subsection{Read and Write Issue Controllers}

Figure ~\ref{fig:eGPU_read_ctl} shows a block diagram of the read access controller. The write access controller has a similar structure, although the request information between the controllers is different. A read request contains an address (typically 16 bits) and a destination register (typically 5 bits), while the write request contains an address (again, typically 16 bits), and data (32 bits). Read and write operations can also be running at the same time; while a read operation will pause the instruction fetch and decode, a write operation can allow the pipeline to continue after it has issued, including a following read operation. There are two types of writes: a blocking write, which holds the instruction pipeline, and a non-blocking write, which does not. A blocking write is used if the same data will likely be used immediately, such as the reordering of data between passes of an FFT, as we will benchmark later in this paper.

Upon receiving a read instruction, the read access controller calculates the number of bank conflicts for each operation. The lower 4 bits of each of the 16 parallel addresses are first converted to a one-hot vector; each vector forms a row of a 2D matrix that indicates which bank that address accesses. We input each column of this matrix into a population counter, which returns a 5-bit value of the number of accesses into each bank. We sort all 16 bank access counts to find the maximum - the number of clock cycles required to complete the current operation is equal to the highest number of bank conflicts. This is repeated, one per clock cycle, for all operations in the instruction. This count value, along with the request information (destination register and addresses) is stored in a circular buffer. The controller will then issue the operations to the shared memory continuously, spaced by the number of clocks required to resolve the bank conflicts. There is a 5 cycle initial latency between receiving the read instruction and issuing the first read operation to the memory, which is the time required to calculate the first set of bank conflicts - this can be seen in the sort network pipeline of figure~\ref{fig:eGPU_read_ctl}. As the number of actual cycles to perform a memory instruction can require hundreds of clocks, this initial latency only has a minor impact on the performance, which is hopefully more than offset by the increased bandwidth of the banked memory. As an example, the 4096-point, Radix-16 FFT used in this work uses 256 threads. A read or write instruction will require 16 to 256 clock cycles, depending on bank conflicts.  

\subsection{Shared Memory Arbitration}

Read or write operations are then issued by the respective controller into the banked memory. The bank conflicts are calculated again, although this time it is to find the sequence of conflicts into any given bank. As with the controller, the 4 LSBs of each of the 16 addresses are converted to one-hot values, creating a 2D matrix of bank accesses. This is the same matrix as calculated in the controller, but it is much less expensive to re-calculate these bits than to buffer and transmit them from the previous block. These bits - which contain a map of the accesses to the respective bank by all 16 lanes - are input to a set of 16 arbiters. On any given clock cycle (bit position of a row), there will be only one mapping from any individual memory bank to any individual lane {\em{i.e}} SP.

Each bank has its own arbiter. The input vector will contain some number of '1's and some number of '0's, although all '1' (this bank is used by all lanes, {\em{i.e.}} maximal bank conflict) and all '0's (this bank is not used at all in this read) are possible. Every read cycle (16 parallel reads) in an instruction (which may contain up to 256 read cycles) may have a different bank conflict mapping, and this may require a varying number of clocks to resolve the bank conflicts. The arbiter will output a mux control signal (in this case a one-hot 16-bit value) to map one of the 16 input addresses to the read address port of one of the memory banks. 

The memory bank outputs must then be mapped to the correct lane, which is implemented with 16 parallel 16-to-1 muxes. The control signals for these muxes are also one-hot, and generated as follows. The input mux control mappings (the 16 arbiter outputs) are delayed by the 3 clock latency of the memory banks, and then transposed. The rows of this matrix are the control signals of each output mux. The logical OR of any column is the writeback signal into the respective SP. The shared memory architecture is depicted in figure~\ref{fig:eGPU_read_arbitrate}.

The write arbitration is very similar, although it is only on the input side of the memory banks. A set of 16, 16-to-1 32-bit data muxes uses the same control as the address muxes. Both data and address muxes in our circuit are one-hot and have a 3-stage pipeline.

A bank mapping example is shown in figure~\ref{fig:eGPU_bankmap}. Although the eGPU has a 16 lane (16 SP) architecture, typically connected to a 16 bank memory, an 8 SP (8 lane) system, connected to an 8 banked memory is described for simplicity. This bank mapping is based on the 3 LSBs of the addresses. From left to right, lane 0 maps to bank 0, lane 1 maps to bank 1, lane 2 also maps to bank 2, and so on. There are bank conflicts to bank 1 which has 3 accesses, and bank 3, which has 2 accesses. If there is any bank with more than one access, then there must be a bank with zero accesses. The bank addresses are each converted to a one-hot 8 element vector, which is assembled to create a 2D matrix of bank addresses. The columns of this matrix describe the individual mappings of lanes to banks. Bank 0 is accessed by lane 0 only. Bank 1 is accessed by lanes 1, 2, and 4. Bank 2 is not accessed in this operation at all. In this example, the maximum number of bank conflicts is 3 to lane 1, and the controller waits 3 cycles before issuing the next operation. (The controller calculated the number of bank conflicts when it received the operation, and stored this value alongside the request information in its buffer).

\subsubsection{8 and 4 Bank Memories}

Reducing the number of banks will reduce the control logic proportionally. The number of ports is the same, so the interface between the processor core and memory is identical. 

From figure~\ref{fig:eGPU_read_arbitrate}, there are now 8 (instead of 16) arbiters, although each arbiter still controls a 16-to-1 mux, to map the 16 input addresses to one of the 8 bank memories. The output mapping is constructed in a similar way. We can see that this essentially cuts the logic and routing requirements of the 8 bank case in half, compared to the 16 bank case. The 4 bank architecture performs memory mapping based on the 2 LSBs of the addresses, and we can follow the impact of this based on what we observed in the 8 bank structure.

\subsubsection{Other Bank Mappings}

The simplest bank mapping is just to use the LSBs of the address for the map value. Other patterns can easily be applied on an instance by instance basis in the FPGA. For complex numbers (the I and Q components are typically stored in adjacent addresses), using a shifted index map (for a 16 bank system, this would use address bits [4:2] rather than [3:0]) can provide significant performance advantages. These are reported as the \textbf{Offset} maps in tables~\ref{tab:results_transpose} and~\ref{tab:results_fft}.

\subsection{Carry Based Arbitration}

\begin{figure}
    \centering
    \includegraphics[scale=0.50]{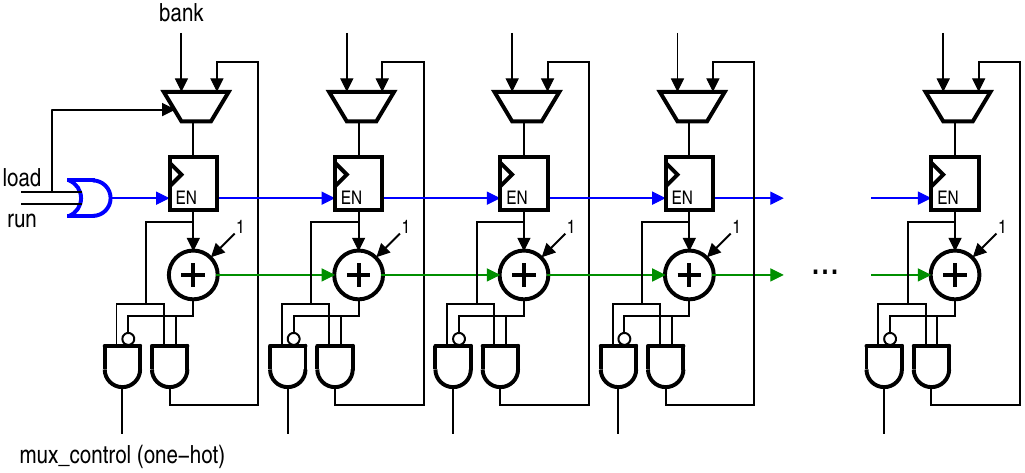}
   \caption{Arbiter Circuit}
   \label{fig:eGPU_arbiter}
\end{figure}

The read and write bank arbitration circuits are both centered around an arbiter circuit, which generates the bank mux controls. The maximum number of active mux operations for any arbiter is known, and the arbiter must complete in this number of clocks. We will now describe a highly efficient carry-chain based arbiter , which is both low cost and high performance, and maps directly to FPGA architectures. This circuit is shown in figure~\ref{fig:eGPU_arbiter}.

A vector defining the accesses to that bank is initially loaded - a '1' bit denotes that the bank the arbiter controls will be used by that lane. All lanes have equal priority and the arbiter starts with the rightmost lane. At each iteration, the value one is subtracted ({\em{i.e.}'-1'}) from the current arbiter value. This flips the next indicated lane marker to '0'. But - it also flips all the previous lane markers back to '1'. (All of the unprocessed lane markers remain unchanged). To correct the re-assertion errors, we zero any location that exhibits a '0' to '1' transition. We also output a '1' at the point of the '1' to '0' transition, which is the current active lane. 

\begin{figure}
    \centering
    \includegraphics[scale=0.80]{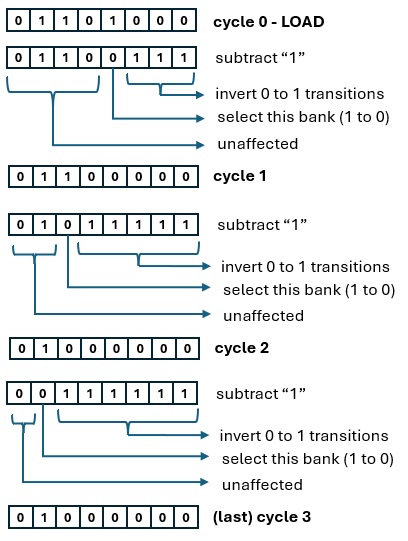}
   \caption{Arbitrate Example}
   \label{fig:eGPU_arbitrate_example}
\end{figure}

Figure~\ref{fig:eGPU_arbitrate_example} shows the state of the arbiter running for Bank 1 of Figure~\ref{fig:eGPU_bankmap}. The '1' to '0' transitions indicate which one-hot mux is selected during that clock cycle.

\section{FPGA Fitting and Results}

\begin{figure}
    \centering
    \includegraphics[scale=0.80]{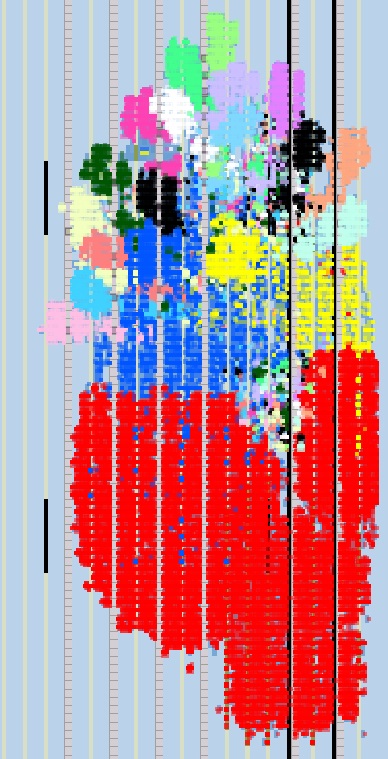}
   \caption{Processor with 16 Bank Shared Memory - the Banked Memory is shown in red.}
   \label{fig:eGPU_16banks}
\end{figure}

Figure~\ref{fig:eGPU_16banks} shows our SIMT Processor with a 16 bank memory (this is the design detailed in table~\ref{tab:bank_resources}). This design closes timing at 771 MHz, without any timing or placement constraints. We were able to reproduce this performance in any Agilex-1 speedgrade device -  performance is limited by the DSP Blocks in FP32 mode. The unrestricted FMax {\em{i.e.}} the critical path outside of the DSP Block is 775 MHz, which is found inside the shared memory.

The shared memory (colored red) is grouped in an almost contiguous block, and the two memory access controllers (the read controller is yellow and the write controller is blue) are placed adjacent to it, with the 16 SPs wrapped around them. The smaller memories (8 and 4 banks) have much faster critical paths, and always close timing around 800 MHz, although the SIMT processor that they are attached to would still be limited by the floating point DSP Blocks to 771 MHz.

\begin{table*}
  \begin{center}
    \caption{SIMT Processor with Various Memory Banks and Architectures}
    \label{tab:bank_resources}
    \begin{tabular}{|c|c||c|c|c|c|c|c|c||}
\hline
  \textbf{Banks} & \textbf{Module} & \textbf{No.} & \textbf{Submod} & \textbf{No.}  & \textbf{ALMs} & \textbf{Regs} & \textbf{M20K} & \textbf{DSP} \\
      \hline\hline
      \multirow{2}{*}{Common} & {SP} & {16} & {-} & {-} & {430} & {1100} & {2} & {2}   \\
    \ & {Fetch/Decode} & {1} & {-} & {-} & {233} & {508} & {2} & {0}   \\
    \hline
    \multirow{5}{*}{4 Banks} & {Read Ctl.} & {1} & {-} & {-} & {342} & {1105} & {6} & {0}   \\
    \ & {Write Ctl.} & {1} & {-} & {-} & {811} & {3114} & {19} & {0}   \\
    \ & {Shared Mem.} & {1} & {-} & {-} & {3225} & {10389} & {32} & {0}   \\
    \ & {} & {} & {Read Arb.} & {4} & {135} & {372} & {0} & {0}   \\
    \ & {} & {} & {Write Arb.} & {4} & {441} & {1166} & {0} & {0}   \\
    \ & {} & {} & {Output Mux} & {16} & {40} & {118} & {0} & {0}   \\
    \hline
    \multirow{5}{*}{8 Banks} &  {Read Control} & {1} & {-} & {-} & {511} & {1595} & {7} & {0}   \\
    \ & {Write Ctl.} & {1} & {-} & {-} & {1094} & {4072} & {19} & {0}   \\
    \ & {Shared Mem.} & {1} & {-} & {-} & {6526} & {20324} & {64} & {0}   \\
    \ & {} & {} & {Read Arb.} & {8} & {145} & {384} & {0} & {0}   \\
    \ & {} & {} & {Write Arb.} & {8} & {448} & {1165} & {0} & {0}   \\
    \ & {} & {} & {Output Mux} & {16} & {80} & {188} & {0} & {0}   \\
    \hline
    \multirow{5}{*}{16 Banks} & {Read Ctl.} & {1} & {-} & {-} & {789} & {2151} & {7} & {0}   \\
    \ & {Write Ctl.} & {1} & {-} & {-} & {1507} & {5245} & {20} & {0}   \\
    \ & {Shared Mem.} & {1} & {-} & {-} & {13105} & {39805} & {128} & {0}   \\
    \ & {} & {} & {Read Arb.} & {16} & {138} & {369} & {0} & {0}   \\
    \ & {} & {} & {Write Arb.} & {16} & {438} & {1164} & {0} & {0}   \\
    \ & {} & {} & {Output Mux} & {16} & {173} & {353} & {0} & {0}   \\
    \hline
    \ {Multi-Port} & {R/W Control} & {1} & {-} & {-} & {700} & {795} & {0} & {0}   \\
    \ {4R-1W} & {Shared Mem.} & {1} & {-} & {-} & {131} & {237} & {64} & {0}   \\
    \hline
    \hline
    \end{tabular}
  \end{center}
\end{table*}

The logic area of the read and write access controllers varies linearly with the number of banks, although the number of M20Ks (used to buffer the read and write requests) is relatively constant. The individual read and write arbitrate cores always use about the same amount of logic, but the number of these cores required is the same as the number of banks. The number of arbitration circuits and the output muxes comprise about 90\% of the logic of the bank memory resources. 

The processors of table~\ref{tab:bank_resources} are configured so that the number of M20Ks is approximately 70\% of the number available by ratio to the shared memory logic (largely the arbitrators and muxes). (In an Agilex-7 device, this is about 70 ALMs to each M20K). We then also increased the M20K to logic ratio, first to 100\%, and finally to 110\% - the speed remained over 771 MHz in all cases for an unconstrained compile. 

The multi-port memory was configured with 64 M20Ks for this example, but the logic size is essentially constant with varying memory sizes.

\subsection{True Cost of a Processor}

To understand which processor memory variant is best, we need to understand the true cost. While table~\ref{tab:bank_resources} shows individual resources by type, the placed cost may be different, especially as the device becomes full. We define our area methodology as follows: (i) memories are node locked to sectors - this will effectively ensure area and performance, despite potentially long routes, even if the processor is incorporated into a larger, dense system design, and (ii) the SPs, instruction fetch and decode, as well as the memory access circuitry are left to place and route in an unconstrained manner. We then express the total area in sector equivalents (16640 ALMs - this gives an accurate footprint, as in the unconstrained placement region the ALMs dominate). 

We were able to achieve our target speed of 771 MHz for all processors (600 MHz for the 4R-2W variant, which configured the M20K memories in the slower emulated true-dual port mode) in a completely unconstrained compile. The 16 bank memory needs about 13K ALMs by itself, and the cost including the read and write controllers is twice that of the SIMT core. In contrast, the multi-port memory (4R-1W, 4R-2W) requires less than 1K ALMs in an unconstrained placement, although the memory is much less efficient (multiple identical copies are required).

\begin{figure}
    \centering
    \includegraphics[scale=0.48]{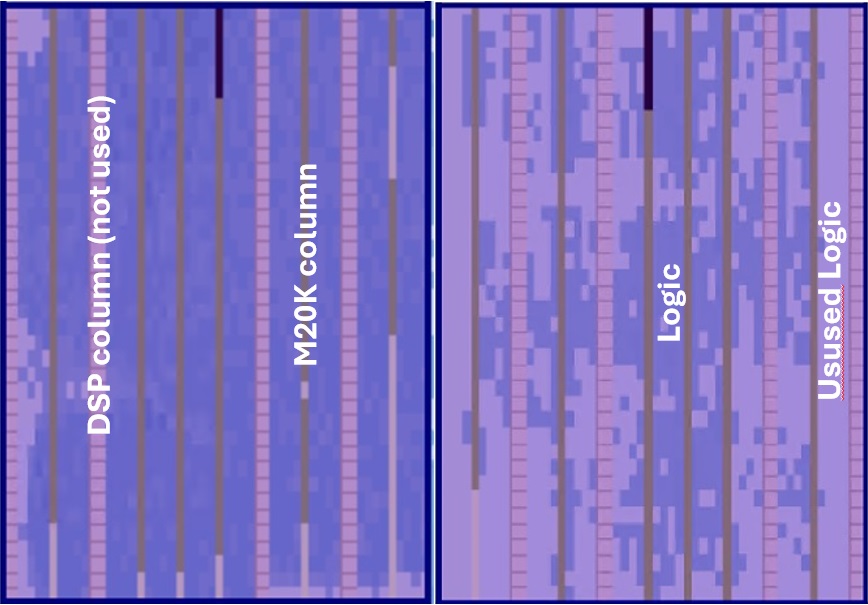}
   \caption{Banked (448KB) vs. Multi-Port (112KB) Memory Placement}
   \label{fig:MemVsMem}
\end{figure}

We node locked a 16 bank, 448 KB shared memory to a single sector, which required 224 M20Ks (shown in the left side of figure~\ref{fig:MemVsMem}). Utilized M20K memories are depicted in dark gray, used ALM logic in dark purple, and unused ALM logic in light purple. This memory used 80\% of the ALMs and 95\% of the M20Ks available. The design closed timing at 738 MHz, although we had to split each memory bank into two, with the upper address bit selecting a half bank. The two additional latency cycles introduced had no material impact on the benchmark performance of the processor. The cost of a 16 bank memory (maximum of 448 KB) is therefore one sector, or 16640 ALMs. As the 8 and 4 bank memories use the identical components to the 16 bank variant, we can state that the cost of these are a 1/2 or 1/4 sector, respectively. The rest of the processor (memory access controllers, SPs, and the instruction fetch and decode) are attached to the memory sector, and placed without any constraints.

The simpler multi-port memory, although appearing to have a smaller logic area, is in fact the same area for larger memories. We node locked a large 112 KB 4W-1W memory into a sector, and needed to add a considerable amount of pipelining to maintain speed. This is because the individual 112 KB memories (there are 4 of these in the 4R architecture) span a large area (requiring 1.5 M20K columns). See the right hand side of figure~\ref{fig:MemVsMem}. Much smaller multi-port memories may require no additional pipelining, and mid-sized examples some amount of pipelining. For simplicity, we will assume that a 64KB (or smaller) memory would require no additional logic, and there would be a linear increase in pipelining required up to a full sector of memory.

\section{Results}

\begin{table*}
\scriptsize
  \begin{center}
    \caption{Transpose Profiling - Different Memory Architectures}
    \label{tab:results_transpose}
    \begin{tabular}{|c|c||c|c|c|c|c|c|c|c||}
\hline
  \textbf{Matrix} & \textbf{Type} & \textbf{4R-1W} & \textbf{4R-2W} & \textbf{16 Banks} & \textbf{16 Banks} & \textbf{8 Banks} & \textbf{8 Banks} & \textbf{4 Banks} & \textbf{4 Banks} \\
  \ & & & & & \textbf{Offset} & & \textbf{Offset} & & \textbf{Offset}\\
      \hline\hline
      \ \multirow{5}{*}{32x32} & {Common Ops} & \multicolumn{8}{c||}{INT OPs : 256,  Immediate OPs : 129,  Other OPs : 6}\\ 
    \ & {Load/Store} & \multicolumn{8}{c||}{64/64}   \\
      \cline{3-10}
    \ & {Load Cycles} & {256} & {256} & {168} & {106} & {290} & {166} & {544} & {288} \\
    \ & {Store Cycles} & {1024} & {512} & {1054} & {1050} & {1048} & {1048} & {1046} & {1046}\\
    \cline {2 -10}
    \ & {Total} & {1671} & {1159} & {1613} & {1547} & {1729} & {1605} & {1981} & {1725}\\
    \ & {Time (us)} & {2.17} & {1.93} & {2.09} & {2.01}  & {2.24}  & {2.08} & {2.57} & {2.24}\\
    \ & {R Bank Eff. (\%)} & {-} & {-} & {38.1} & {60.4} & {22.1}  & {38.6}  & {11.8} & {22.2} \\
    \ & {W Bank Eff. (\%)} & {-} & {-} & {6.1} & {6.1} & {6.1}  & {6.1}  & {6.1} & {6.1} \\
    \hline
    
    \ \multirow{5}{*}{64x64} & {Common Ops} & \multicolumn{8}{c||}{INT OPs : 192,  Immediate OPs : 161,  Other OPs : 6}\\ 
    \ & {Load/Store} & \multicolumn{8}{c||}{256/256}   \\
      \cline{3-10}
    \ & {Load Cycles} & {1024} & {1024} & {1184} & {672} & {2184} & {1160} & {4224} & {2176} \\
    \ & {Store Cycles} & {4096} & {2048} & {4216} & {4200} & {4192} & {4192} & {4184} & {4184}\\
    \cline {2 -10}
    \ & {Total} & {5479} & {3431} & {5759} & {5231} & {6735} & {5711} & {8767} & {6719}\\
    \ & {Time (us)} & {7.1} & {5.72} & {7.46} & {6.78}  & {8.74}  & {7.41} & {11.37} & {8.71}\\
    \ & {R Bank Eff. (\%)} & {-} & {-} & {21.6} & {38.9} & {11.7}  & {22.1}  & {6.1} & {11.8} \\
    \ & {W Bank Eff. (\%)} & {-} & {-} & {6.1} & {6.1} & {6.1}  & {6.1}  & {6.1} & {6.1} \\
    \hline
    
    \ \multirow{5}{*}{128x128} & {Common Ops} & \multicolumn{8}{c||}{INT OPs : 160,  Immediate OPs : 129,  Other OPs : 6}\\ 
    \ & {Load/Store} & \multicolumn{8}{c||}{1024/1024}   \\
      \cline{3-10}
    \ & {Load Cycles} & {4096} & {4096} & {8832} & {4672} & {16928} & {8736} & {16896} & {16896} \\
    \ & {Store Cycles} & {16384} & {8192} & {16864} & {16800} & {16768} & {16768} & {16736} & {16736}\\
    \cline {2 -10}
    \ & {Total} & {20775} & {12583} & {25991} & {21767} & {33991} & {25799} & {34017} & {34017}\\
    \ & {Time (us)} & {26.95} & {20.97} & {33.71} & {28.23}  & {44.09}  & {33.46} & {44.12} & {44.12}\\
    \ & {R Bank Eff. (\%)} & {-} & {-} & {11.6} & {21.9} & {6.0}  & {11.7}  & {6.1} & {6.1} \\
    \ & {W Bank Eff. (\%)} & {-} & {-} & {6.1} & {6.1} & {6.1}  & {6.1}  & {6.1} & {6.1} \\
    \hline

    \end{tabular}
  \end{center}
\end{table*}

\begin{table*}
\scriptsize
  \begin{center}
    \caption{FFT Profiling - Different Memory Architectures}
    \label{tab:results_fft}
    \begin{tabular}{|c|c||c|c|c|c|c|c|c|c|c||}
\hline
  \textbf{Points} & \textbf{Type} & \textbf{4R-1W} & \textbf{4R-2W} & \textbf{4R-1W-VB}  & \textbf{16 Banks} & \textbf{16 Banks} & \textbf{8 Banks} & \textbf{8 Banks} & \textbf{4 Banks} & \textbf{4 Banks} \\
  \ & & & & & & \textbf{Offset} & & \textbf{Offset} & & \textbf{Offset}\\
      \hline\hline
      \ \multirow{5}{*}{Radix 4} & {Common Ops} & \multicolumn{9}{c||}{FP OPs : 13440, INT OPs : 2880,  Immediate OPs : 1287,  Other OPs : 244}\\ 
    \ & {D Load/Store} & \multicolumn{9}{c||}{3072/3072}   \\
    \ & {TW Load} & \multicolumn{9}{c||}{1920}   \\
      \cline{3-11}
    \ & {D Load Cycles} & {12228} & {12228} & {12228} & {11200} & {7104} & {19248}  & {11120} & {29440} & {19200} \\
    \ & {W Load Cycles} & {7680} & {7680} & {7680} & {24152} & {21548} & {27134}  & {24070} & {29152} & {27104} \\
    \ & {Store Cycles} & {49152} & {24576} & {24576} & {10960} & {6864} & {19008}  & {10880} & {29200} & {18960}\\
    \cline {2 -11}
    \ & {Total} & {86817} & {62214} & {62214} & {64063} & {53267} & {80361}  & {63821} & {105543} & {82915}\\
    \ & {Time (us)} & {112.60} & {103.74} & {80.69} & {83.09} & {69.09} & {104.23} & {82.78} & {136.89} & {107.54}\\
    \ & {Efficiency (\%)} & {15.5} & {21.6} & {21.6} & {21.0} & {25.2}  & {16.7}  & {21.1} & {12.7} & {16.2}\\
    \ & {D Bank Eff. (\%)} & {-} & {-} & - & {28.0} & {44.8}  & {16.2}  & {28.2} & {10.5} & {16.2}\\
    \ & {TW Bank Eff (\%)} & {-} & {-} & - & {7.9} & {8.9}  & {7.1}  & {8.0} & {6.6} & {7.1}\\
    \hline

    \ \multirow{5}{*}{Radix 8} & {Common Ops} & \multicolumn{9}{c||}{FP OPs : 11840, INT OPs : 3456,  Immediate OPs : 523,  Other OPs : 108}\\ 
    \ & {D Load/Store} & \multicolumn{9}{c||}{2048/2048}   \\
    \ & {TW Load} & \multicolumn{9}{c||}{1344}   \\
      \cline{3-11}
    \ & {D Load Cycles} & {8192} & {8192}  & {8192}  & {12624}  & {7425} & {15424} & {12448}  & {21504} & {15320}\\
    \ & {W Load Cycles} & {5376} & {5376} & {5376} & {16712} & {13844} & {18122}  & {16608} & {20128} & {18080} \\
    \ & {Store Cycles}  & {32768} & {16384} & {20480} & {12224} & {7104} & {15104}  & {12128} & {21184} & {15040}\\ 
    \cline {2 -11}
    \ & {Total} & {62263} & {45879} & {49975} & {57487} & {44300} & {64577}  & {57111} & {78743} & {65367}\\
    \ & {Time (us)} & {80.76} & {76.47} & {64.82} & {74.56} & {57.46}  & {83.76}  & {74.07} & {102.13} & {84.78}\\
    \ & {Efficiency(\%)} & {19.0} & {25.8} & {23.7} & {20.6} & {26.7}  & {18.3}  & {20.7} & {15.0} & {18.1}\\
    \ & {D Bank Eff. (\%)} & {-} & {-} & - & {16.8} & {28.8}  & {13.6}  & {16.9} & {9.7} & {13.6}\\ 
    \ & {TW Bank Eff. (\%)} & {-} & {-} & - & {8.0} & {9.7}  & {7.4}  & {8.1} & {6.7} & {7.4}\\ 
    \hline

    \ \multirow{5}{*}{Radix 16} & {Common Ops} & \multicolumn{9}{c||}{FP OPs : 12384, INT OPs : 2192,  Immediate OPs : 276,  Other OPs : 90}\\ 
    \ & {D Load/Store} & \multicolumn{9}{c||}{1536/1536}   \\
    \ & {TW Load} & \multicolumn{9}{c||}{960}   \\
      \cline{3-11}
    \ & {D Load Cycles} & {6144} & {6144}  & {6144}  & {12160}  & {11136} & {13920} & {12000}  & {17920} & {13824}\\
    \ & {W Load Cycles} & {3840} & {3840} & {3840} & {10888} & {9848} & {14876}  & {10780} & {14272} & {12244} \\
    \ & {Store Cycles}  & {24576} & {12228} & {14336} & {11680} & {10652} & {13440}  & {11520} & {17440} & {13344}\\ 
    \cline {2 -11}
    \ & {Total} & {49442} & {37214} & {39262} & {49670} & {46578} & {57177}  & {49242} & {64483} & {54354}\\
    \ & {Time (us)} & {64.13} & {62.02} & {50.92} & {64.53} & {60.41}  & {74.16}  & {63.87} & {83.64} & {70.50}\\
    \ & {Efficiency (\%)} & {25.0} & {33.3} & {31.5} & {24.9} & {26.6}  & {21.7}  & {25.1} & {19.2} & {22.8}\\
    \ & {D Bank Eff. (\%)} & {-} & {-} & - & {13.2} & {14.4}  & {11.4}  & {13.3} & {8.8} & {11.5}\\
    \ & {TW Bank Eff. (\%)} & {-} & {-} & - & {8.8} & {9.7}  & {6.4}  & {8.9} & {6.7} & {7.8}\\
    \hline
    \end{tabular}
  \end{center}
\end{table*}

\begin{figure*}
    \centering
    \includegraphics[scale=0.90]{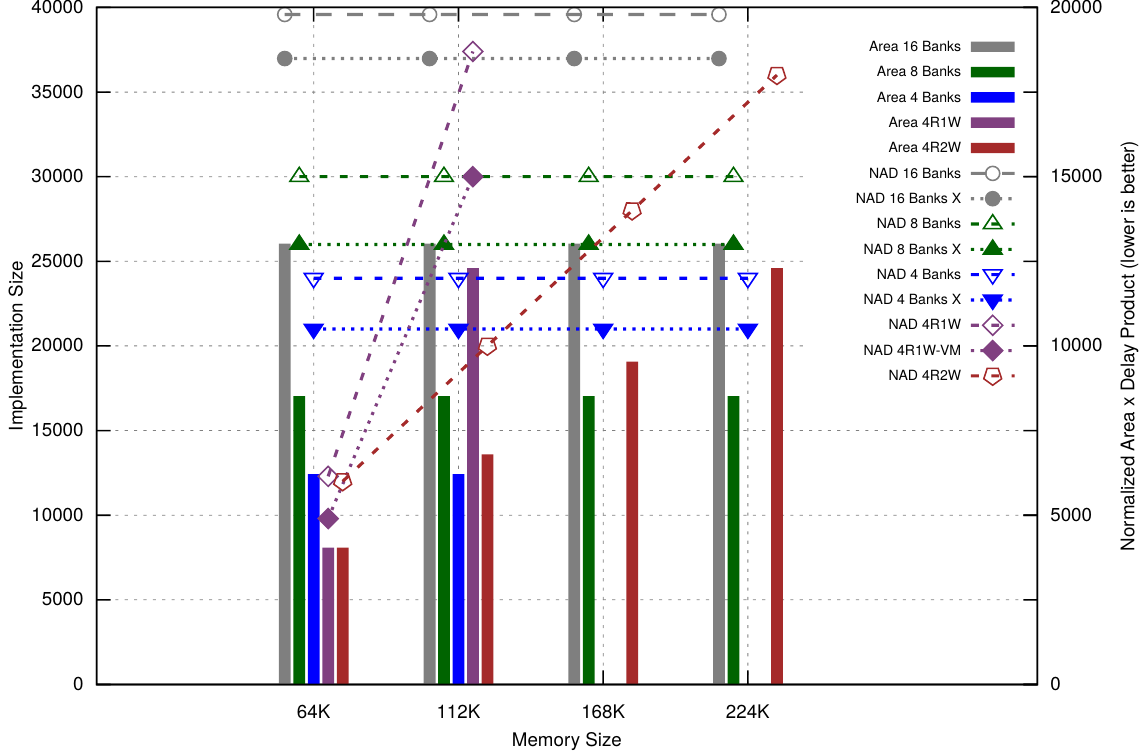}
   \caption{Cost vs. Performance (lower is better)}
   \label{fig:MemCost}
\end{figure*}

We ran a total of 52 testcase combinations to examine the impact of different types of shared memory architectures. All benchmarks were written in assembler.

We first compared the transpose of three matrices (32x32, 64x64, and 128x128) over 8 memory types. The results can be found in Table~\ref{tab:results_transpose}. The first two used multi-port 4R (4-read port) multi-port memory, with 1W (single write port) and 2W (two write port) interfaces. The multi-port memory based architectures were marginally faster than the 16 banked memory ones. Data read cycles were significantly more efficient than the writes with banked memory as the reads are across columns (which would naturally be mapped in different banks), and the writes are down columns, where individual columns might well be mapped to a single bank. The write efficiencies are all $\approx$ 6\%, which would correlate to a 1:16 access ratio - in other words, any given writeback of the transposed data is into a single bank. On the read side, the highest efficiency is 60.4\% for the 32x32 16 bank (offset) core. This is actually 100\% efficiency at the shared memory - the gap is because of the latency through the read access control and writeback to the SPs. The complex bank mapping improves the performance of the transpose benchmarks by about 10\%, despite the matrix containing only real numbers. This can be explained by the interaction between the location of the matrix elements in the banks and the access patterns of the reads.

We ran a total of 27 FFT combinations, with radix 4, 8, and 16 over 9 different memory architectures. The 4096-point FFT provides a large dataset (nearly 64KB with the required twiddle coefficients), with multiple whole dataset accesses. The FFT exercises a SIMT memory architecture thoroughly, with large dataset transfers with changing memory access patterns. We also have a detailed prior work for comparison~\cite{HEART_FFT}. Although many GPGPU FFTs use constant geometry FFT algorithms like Pease~\cite{PeaseFFT} or Stockham~\cite{StockhamFFT}, we program our FFTs using the standard Cooley-Tukey~\cite{Cooley-Tukey} algorithm, as our goal is to compare the effect of the different memory architecture, not necessarily to achieve maximum performance. 

Table~\ref{tab:results_fft} shows the different FFT results. One of the multi-port memory types (4R-1W-VM) has an additional instruction that allows 4W to be issued. The effect of this is that the multi-port memory becomes 4 separate memories for that dataset - this can be used to increase bandwidth for algorithms like the FFT. The explanation of this mode is beyond the scope of this paper, but the effect is to improve write bandwidth on average to that of the 4R-2W memory, but at the higher system speed of 771 MHz. All other FFTs use the same code for a given radix.

The 16 bank memory, with the complex bank mapping, typically gives us the highest performance. We also show efficiency (the percentage of time that the core is calculating the FFT, which does not include address generation or shared memory accesses). The efficiency of our processor is up to 33\% for the multi-port memory version (27\% for the banked memory version), which compares favorably with the latest cuFFT results of 33\% for the A100~\cite{A100} for a 4096-point FFT~\cite{cufft}. 

\section{What is the best Memory Type for soft SIMT Processors?}

Figure~\ref{fig:MemCost} graphs the cost of the SIMT core with different memory architectures, for 64KB, 112KB, 168KB, and 224KB. First, absolute area (the vertical bars) is expressed in ALM footprints (the other resources, the M20K memories and DSP Blocks, reside within the perimeter defined by that footprint, as described in section IV.A). 
The banked cores have a constant area, with the 16 banked memory taking up an entire sector, the 8 banked memory half a sector, and the 4 bank memory 1/4 a sector. In contrast, the multi-port cores increase in area with shared memory size, because of the additional pipelining required (as shown in Fig.~\ref{fig:MemVsMem}). If we take a full sector as the maximum allowed size for the shared memory (which is twice the cost of the rest of the processor), the multi-port memories reach the roofline quickly. The 4R-1W memory can support a maximum of only 112 KB, and the 4R-2W memory (which uses the M20K blocks in the quad-port mode) 224KB.

We use the radix-16 (4096 points) FFT as our performance benchmark, which requires both high memory bandwidth and floating point operations, so is representative of expected SIMT applications. The dashed lines represent normalized performance (lower is better) compared to the slowest core. The multi-port memories are more efficient for smaller dataset sizes ({\em{e.g.}} 256 points), but quickly become more expensive than some of the banked memory based cores. The smaller banked memories are more efficient (performance per unit area) than the larger banked memories, but more banks mean more absolute performance, as shown in table~\ref{tab:results_fft}. If larger dataset sizes are needed, the most efficient way to do this is to add ports - there is no point in increasing the memory size of the 4 bank memory beyond 112KB, as the additional footprint would contain the logic to support more banks. 

The 4096-point FFT requires 64KB (data and twiddles), so all the memory sizes in figure~\ref{fig:MemCost} would be able to accommodate it. Larger memory sizes would be needed for multi-batch cases (each additional dataset needs 32KB), or if several different programs were run.

\section{Conclusions}

The many-ported banked memory architectures are not always the best choice for high-performance SIMT processors. Multi-port memories typically have a higher performance on the benchmarks we evaluated. The multi-port memory based processors can also be significantly smaller, but only for smaller datasets - up to about 64KB of shared memory. Larger multi-port memories require a significant amount of real estate, as an increasing amount of pipelining and other logic is required to maintain speed, especially for designs with constrained fitting. Banked memory architectures have a relatively constant area, whether constrained placement is used or not, even for larger shared memories approaching 500KB.

By aligning the logic depth and pipeline structure of the banked memory to the FPGA architecture, we are able to achieve and maintain a high performance (over 735 MHz for a tightly constrained placement and 771 MHz for an unconstrained placement) SIMT processor. The multi-port based processor also closes timing at these speeds. The best choice of shared memory architecture is then most likely determined by the dataset size: applications that require less shared memory should use the multi-ported memories, larger data requirements should use banked memory. The choice between the two types of memory will also be influenced by memory access patterns - we have used the FFT here as a reasonable representative SIMT application. The one advantage of the FPGA is that we will be able to change our memory architecture to suit our particular design. 

Despite using only a vanilla SIMT core with straightforward implementations of algorithms, our efficiency is about the same as the Nvidia A100 for FFT. The advantage of the FPGA is flexibility - we have the option of adding hardware features (such as a complex FP ALU), additional memory types, adjusting the shared memory size to contain other coefficient configurations that will be more efficient to read for the banking selected, as well as varying the bank mapping - all to improve the final performance of our SIMT processor.  

\bibliographystyle{IEEEtran}
\bibliography{references.bib}

\end{document}